\documentclass[]{sig-alternate}

\usepackage{amsmath,amssymb}
\usepackage{alltt,mathrsfs}
\usepackage{multirow,rotating}

\usepackage{colortbl}
\definecolor{kugray5}{RGB}{224,224,224}

\def\val {\ensuremath{\mathrm{val}}}

\def\partial{\delta}

\usepackage[plainpages=false,pdfpagelabels,colorlinks=true,citecolor=blue,hypertexnames=false]{hyperref}

\usepackage{comment}
\makeatletter
\newif\if@restonecol
\makeatother

\usepackage[lined,boxed,ruled]{algorithm2e}
\SetKwInOut{Input}{input}
\SetKwInOut{Output}{output}
\def\RDAC{\mathsf{RDAC}}
\def\DAC{\mathsf{DAC}}
\def\LinAlg{\mathsf{LinSolve}}
\def\Sylvester{\mathsf{Sylvester}}
\def\NewtonAE{\mathsf{NewtonAE}}
\def\DiffSylvester{\mathsf{DiffSylvester}}
\def\DiffSylvesterDifferential{\mathsf{DiffSylvesterDifferential}}
\def\PolCoeffsDE{\mathsf{PolCoeffsDE}}

\def\Spec{\operatorname{Spec}}

\def\N {\ensuremath{\mathbb{N}}}

\def\K {\ensuremath{\mathbb{K}}}

\def\M {\ensuremath{\mathsf{M}}}
\def\I {\ensuremath{\int_q}}
\def\Id{\ensuremath{\mathsf{Id}}}

\def\mA {\ensuremath{{A}}}
\def\mB {\ensuremath{{B}}}
\def\mC {\ensuremath{{C}}}
\def\mD {\ensuremath{{D}}}

\def\mF {\ensuremath{{F}}}

\def\mP {\ensuremath{{P}}}

\def\mR {\ensuremath{{R}}}

\def\mU {\ensuremath{{U}}}
\def\mV {\ensuremath{{V}}}
\def\mW {\ensuremath{{W}}}
\def\mY {\ensuremath{{Y}}}

\def\mzero {\ensuremath{{0}}}

\newtheorem{Def}{Definition}

\newtheorem{Theo}{Theorem}
\newtheorem{Prop}{Proposition}
\newtheorem{Lemma}{Lemma}

\begin{document}

% Start of patch for our 5 authors - cf FAQ on ACM webpage
\def\more-auths{%
\end{tabular}
\begin{tabular}{c}}
% End of patch
\title{Power Series Solutions of\\ Singular (q)-Differential Equations}
\numberofauthors{3}
\author{
\alignauthor Alin Bostan\\[2mm]Bruno Salvy\\[2mm]
	\affaddr{Algorithms Project}\\
	\affaddr{Inria (France)}\\
	\email{Alin.Bostan@inria.fr}\\
	\email{Bruno.Salvy@inria.fr}
\alignauthor Muhammad F. I. Chowdhury\\[2mm]
          \'Eric Schost\\[2mm]
	\affaddr{ORCCA and CS Department}\\
	\affaddr{University of Western Ontario}\\
	\affaddr{London, ON (Canada)}\\
        \email{mchowdh3@csd.uwo.ca}\\
	\email{eschost@uwo.ca}
\alignauthor Romain Lebreton\\[2mm]
	\affaddr{\'Equipe MAX}\\
	\affaddr{LIX, \'Ecole polytechnique}\\
	\affaddr{Palaiseau (France)}\\
	\email{lebreton@lix.polytechnique.fr}
%% \and	
%% \alignauthor Bruno Salvy\\
%% 	\affaddr{Algorithms Project}\\
%% 	\affaddr{Inria}\\
%% 	\affaddr{France}\\
%% 	\email{Bruno.Salvy@inria.fr}
%% \alignauthor \'Eric Schost\\
%% 	\affaddr{ORCCA and CS Department}\\
%% 	\affaddr{University of Western Ontario}\\
%% 	\affaddr{London, ON, Canada}\\
%% 	\email{eschost@uwo.ca}
}

\maketitle
\begin{abstract}
We provide algorithms computing power series solutions of a large
class of differential or $q$-differential equations or systems. Their number of
arithmetic operations grows linearly with the precision, up to
logarithmic terms.
\end{abstract}

%%%%%%%%%%%%%%%%%%%%%%%%%%%%%%%%%%%%%%%%%%%%%%%%%%%%%%%%%%%%
%%%%%%%%%%%%%%%%%%%%%%%%%%%%%%%%%%%%%%%%%%%%%%%%%%%%%%%%%%%%
%%%%%%%%%%%%%%%%%%%%%%%%%%%%%%%%%%%%%%%%%%%%%%%%%%%%%%%%%%%%

\section{Introduction}
Truncated power series are a fundamental class of objects of computer
algebra. Fast algorithms are known for a large number of operations
starting from addition, derivative, integral and product and extending to
quotient, powering and several more. The main open problem is
composition: given two power series $f$ and~$g$, with~$g(0)=0$, known
mod~$x^N$, the best known algorithm computing~$f(g) \bmod x^N$ has
a cost which is roughly that of~$\sqrt{N}$ products in precision $N$;
it is not known whether quasi-linear (i.e., linear up to logarithmic factors) complexity is possible in
general. Better results are known over finite
fields~\cite{Bernstein1998,KedlayaUmans2011} or when more information
on~$f$ or~$g$ is available. Quasi-linear complexity has been reached
when~$g$ is a polynomial~\cite{BrKu78}, an algebraic
series~\cite{Hoeven2002}, or belongs to a large class containing for
instance the expansions of~$\exp(x)-1$
and~$\log(1+x)$~\cite{BostanSalvySchost2008}.

One motivation for this work is to deal with the case when~$f$ is the solution of a
given differential equation. Using the chain rule, a differential
equation for~$f(g)$ can be derived, with coefficients that are power
series. We focus on the case when this equation is linear, since in
many cases linearization is possible~\cite{BoChOlSaScSc07}. When the
order~$n$ of the equation is larger than~1, we use the classical
technique of converting it into a first-order equation over vectors,
so we consider equations of the form
\begin{equation}\label{eq:maindiff}
x^k \partial(\mF) = \mA \mF + \mC,
\end{equation}
where $\mA$ is an $n\times n$ matrix over the power series ring
$\K[[x]]$ ($\K$ being the field of coefficients), $\mC$ and the
unknown $\mF$ are size $n$ vectors over $\K[[x]]$ and for the
moment~$\partial$ denotes the differential operator $d/dx$.  The
exponent $k$ in~\eqref{eq:maindiff} is a non-negative integer that
plays a key role for this equation.

By {\em solving} such equations, we mean computing a vector $\mF$ of
power series such that~\eqref{eq:maindiff} holds modulo $x^N$. For
this, we need only compute $\mF$ polynomial of degree less than $N+1$
(when $k=0$) or $N$ (otherwise). Conversely, when~\eqref{eq:maindiff}
has a power series solution, its first~$N$ coefficients can be computed
by solving~\eqref{eq:maindiff} modulo $x^N$ (when $k\neq0$) or~$x^{N-1}$
(otherwise).

If~$k=0$ and the field~$\K$ has characteristic~0, then a formal Cauchy
theorem holds and~\eqref{eq:maindiff} has a unique vector of power
series solution for any given initial condition. In this situation,
algorithms are known that compute the first $N$ coefficients of the
solution in quasi-linear complexity~\cite{BoChOlSaScSc07}. In this
article, we extend the results of~\cite{BoChOlSaScSc07} in three
directions:

\smallskip\noindent{\bf Singularities.}  We deal with the case
when~$k$ is positive. A typical example is the computation of the
composition~$F=f(g)$ when~$f$ is Gauss' ${}_2F_1$ hypergeometric
series. Although $f$ is a very nice power series
\[f=1+\frac{ab}{c}x+\frac{a(a+1)b(b+1)}{c(c+1)}\frac{x^2}{2!}+\dotsb,\]
we exploit this structure indirectly only. We start from the
differential equation
\begin{equation}\label{eq:hypergeom}
x(x-1)f''+(x(a+b+1)-c)f'+abf=0
\end{equation}
and build up and solve the more complicated
\begin{equation*}\label{eq:hypergeom2}
\frac{g(g-1)}{g'^2}F''+\frac{g'^2(g(a+b+1)-c)+(g-g^2)g''}{g'^3}F'+abF=0
\end{equation*}
in the unknown~$F$, $g$ being given, with
$g(0)=0$. Equation~\eqref{eq:hypergeom} has a leading term that is
divisible by~$x$ so that Cauchy's theorem does not apply and indeed
there does not exist a basis of two power series solutions. This
behavior is inherited by the equation for $F$, so that the techniques
of~\cite{BoChOlSaScSc07} do not apply --- this example is actually
already mentioned in~\cite{BrKu78}, but the issue with the singularity
at $0$ was not addressed there. We show in this article how to
overcome this singular behavior and obtain a quasi-linear complexity.

\smallskip\noindent{\bf Positive characteristic.}  Even when~$k=0$,
Cauchy's theorem does not hold in positive characteristic and
Eq.~\eqref{eq:maindiff} may fail to have a power series solution (a
simple example is~$F'=F$). However, such an equation may have a
solution modulo $x^N$. Efficient algorithms finding such a solution
are useful in conjunction with the Chinese remainder theorem. Other
motivations for considering algorithms that work in positive
characteristic come from applications in number-theory based
cryptology or in
combinatorics~\cite{BostanMorainSalvySchost2008,BostanSalvySchost2008,BostanSchost2009}.

Our objectives in this respect are to overcome the lack of a Cauchy
theorem, or of a formal theory of singular equations, by giving
conditions that ensure the existence of solutions at the required
precisions. More could probably be said regarding the $p$-adic
properties of solutions of such equations (as
in~\cite{BoGoPeSc05,LeSi08}), but this is not the purpose of this
paper.

\smallskip\noindent{\bf Functional Equations.}  The similarity between
algorithms for linear differential equations and for linear difference
equations is nowadays familiar to computer algebraists. We thus use
the standard technique of introducing $\sigma:\K[[x]] \to \K[[x]]$ a
unitary ring morphism and letting $\partial:\K[[x]] \to \K[[x]]$
denote a $\sigma$-derivation, in the sense that $\partial$ is
$\K$-linear and that for all $f,g$ in $\K[[x]]$, we have
$$\partial(fg)=f\partial(g)+\partial(f)\sigma(g).$$ These definitions,
and the above equality, carry over to matrices over $\K[[x]]$. Thus,
our goal is to solve the following generalization
of~\eqref{eq:maindiff}:
\begin{equation}\label{eq:main}
x^k \partial(\mF) = \mA \sigma(\mF) + \mC.
\end{equation}
As above, we are interested in computing a vector $\mF$ of power
series such that~\eqref{eq:main} holds  $\bmod~x^N$.

One motivation for this generalization comes from coding theory. The
list-decoding of the folded Reed-Solomon
codes~\cite{GuruswamiRudra2008} leads to an equation
$Q(x,f(x),f(qx))=0$ where $Q$ is a known polynomial. A linearized
version of this is of the form~\eqref{eq:main}, with
$\sigma:\phi(x)\mapsto\phi(qx)$. In cases of interest we have~$k=1$,
and we work over a finite field.

In view of these applications, we restrict ourselves to the following
setting:
\[\partial(x)=1,\qquad \sigma: x \mapsto qx,\]
for some $q \in \K\setminus\{0\}$. Then, there are only two possibilities:
\begin{itemize}
\item $q=1$ and $\partial:f\mapsto f'$
   ({\em differential} case);
\item $q \ne 1$ and $\partial:f\mapsto\frac{f(qx)-f(x)}{x(q-1)}$ 
  ({\em $q$-differential} case). 
\end{itemize}
As a consequence, $\partial(1)=0$ and for all $i\ge0$, we have
\[\partial(x^i)=\gamma_i x^{i-1}\text{ with }
\gamma_0=0\text{ and }\gamma_i=1+q+\cdots+q^{i-1}\text{ $(i>0)$}.\]
By linearity, given $f=\sum_{i \ge 0} f_i x^i\in\K[[x]]$, 
$$\partial(f)=\sum_{i \ge 1} \gamma_if_i x^{i-1}$$ can be computed
$\bmod~x^N$ in $O(N)$ operations, as can~$\sigma(f)$.  Conversely,
assuming that $\gamma_1,\dots,\gamma_n$ are all non-zero in~$\K$,
given $f$ of degree at most $n-1$ in $\K[x]$, there exists a unique
$g$ of degree at most $n$ such that $\partial(g)=f$ and $g_0=0$; it is
given by $g=\sum_{0 \le i \le n-1} f_i/\gamma_{i+1} x^{i+1}$ and can
be computed in~$O(N)$ operations. We denote it by $g=\I f$. In
particular, our condition excludes cases where $q$ is a root of unity
of low order.

\smallskip\noindent{\bf Notation and complexity model.}  We adopt the
convention that uppercase letters denote matrices or vectors while lowercase
letters denote scalars. The set of $n \times m$ matrices over a ring
$R$ is denoted $\mathscr{M}_{n,m}(R)$; when $n=m$, we write
$\mathscr{M}_{n}(R)$.  If $f$ is in $\K[[x]]$, its degree $i$
coefficient is written $f_i$; this carries over to matrices. The
identity matrix is written $\Id$ (the size will be obvious from
the context). To avoid any confusion, the entry~$(i,j)$ of a
matrix~$M$ is denoted~$M^{(i,j)}$. 

Our algorithms are sometimes stated with input in~$\K[[x]]$, but it is
to be understood that we are given only truncations of~$A$ and~$C$ and
only their first~$N$ coefficients will be used.

The costs of our algorithms are measured by the number of arithmetic
operations in~$\K$ they use. We let $\M: \N \rightarrow \N$ be such
that for any ring $R$, polynomials of degree less than~$n$ in~$R[x]$
can be multiplied in $\M(n)$ arithmetic operations
in~$R$. We assume that $\M(n)$ satisfies the usual
assumptions of~\cite[\S8.3]{GaGe99}; using Fast Fourier Transform,
$\M(n)$ can be taken in $O(n \log (n)\,\log\log
(n))$~\cite{CaKa91,ScSt71}. We note~$\omega\in(2,3]$ a
constant such that two matrices in $\mathscr{M}_n(R)$ can be
multiplied in $O(n^\omega)$ arithmetic operations in~$R$. The current
best  bound is $\omega <
2.3727$~(\cite{VassilevskaWilliams11} following~\cite{CoWi90,Stothers10}).

Our algorithms rely on linear algebra techniques; in particular, we
have to solve several systems of non-homogeneous linear equations. For
$U$ in $\mathscr{M}_n(\K)$ and $V$ in $\mathscr{M}_{n,1}(\K)$, we
denote by $\LinAlg(UX=V)$ a procedure that returns $\bot$ if there is
no solution, or a pair $F,K$, where $F$ is in $\mathscr{M}_{n,1}(\K)$
and satisfies $UF=V$, and $K\in \mathscr{M}_{n,t}(\K)$, for some $t\le
n$, generates the nullspace of $U$. This can be done in time
$O(n^\omega)$. In the pseudo-code, we adopt the convention that if a
subroutine returns $\bot$, the caller returns $\bot$ too (so we do not
explicitly handle this as a special case).

\smallskip\noindent{\bf Main results.}  Equation~\eqref{eq:main} is
linear, non-homogeneous in the coefficients of~$F$, so our output
follows the convention mentioned above. We call {\em generators} of
the solution space of Eq.~\eqref{eq:main} at precision $N$ either
$\bot$ (if no solution exists) or a pair $F,K$ 
where $F\in\mathscr{M}_{n,1}(\K[x])$ and $K\in\mathscr{M}_{n,t}(\K[x])$ with $t \le
nN$, such that for $G\in \mathscr{M}_{n,1}(\K[x])$, with $\deg(G)<N$,
$x^k \partial(G) = A \sigma(G) + C \bmod x^N$ if and only if $G$ can
be written $G=F + K B$ for some $B\in \mathscr{M}_{t,1}(\K)$. 

Seeing Eq.~\eqref{eq:main} as a linear system, one can obtain such an
output using linear algebra in dimension~$nN$. While this solution
always works, we  give algorithms of much better complexity, under
some assumptions related to the spectrum~$\Spec A_0$
of the constant coefficient~$\mA_0$ of~$\mA$. First, we simplify our problem: we 
consider the case $k=0$ as a special case of the case $k=1$. Indeed,
the equation $\partial(\mF) = \mA \sigma(\mF) + \mC \bmod x^N$ is
equivalent to $x\partial(\mF) = P \sigma(\mF) +Q \bmod x^{N+1}$, with
$P=x\mA$ and $Q=x\mC$. Thus, in our results, we only distinguish the
cases $k=1$ and $k>1$.

\begin{Def} The matrix $A_0$ has \emph{good spectrum at precision~$N$} 
when one of the following holds:
\begin{itemize}
\item $k=1$ and $\Spec A_0\cap(q^i\Spec A_0-\gamma_i)=\emptyset$ for $1\le i < N$
\item $k>1$, $A_0$ is invertible and
  \begin{itemize}\setlength{\itemindent}{-10pt}
  \item $q=1$, $\gamma_1,\dots,\gamma_{N-k}$ are non-zero, $|\Spec
    A_0|=n$ and $\Spec A_0\subset\K$;
  \item $q\ne 1$ and $\Spec A_0\cap q^i\Spec A_0=\emptyset$ for $1\le i<N$.
  \end{itemize}
\end{itemize}
\end{Def}
In the classical case when $\K$ has characteristic 0 and $q=1$, if
$k=1$, $A_0$ has good spectrum when no two eigenvalues of $A_0$ differ
by a non-zero integer (this is e.g. the case when $A_0=0$, which is
essentially the situation of Cauchy's theorem; this is also the case
in our~${}_2F_1$ example whenever $c\,\val(g)$ is not an integer,
since $\Spec A_0=\{0,\val(g)(1-c)-1\}$).

These conditions could be slightly relaxed, using gauge
transformations (see~\cite[Ch.~2]{Balser00}
and~\cite{BaBrPf10,BaPf99}).  Also, for $k>1$ and
$q=1$, we could drop the assumption that the eigenvalues are in $\K$,
by replacing $\K$ by a suitable finite extension, but then our
complexity estimates would only hold in terms of number of operations
in this extension.

\smallskip

As in the non-singular case~\cite{BoChOlSaScSc07}, we develop two
approaches. The first one is a divide-and-conquer method. The problem
is first solved at precision~$N/2$ and then the computation at
precision~$N$ is completed by solving another problem of the same type
at precision~$N/2$. This leads us to the following result, proved in
Section~\ref{sec:dac} (see also that section for comparison to
previous work). In all our cost estimates, we consider $k$ constant,
so it is absorbed in the big-Os.
\begin{Theo}\label{theo:1}
Algorithm~\ref{algo:DAC} computes generators of the solution space of Eq.~\eqref{eq:main} at precision
	  $N$ by a divide-and-conquer approach.
  Assuming $A_0$ has good spectrum at precision~$N$, it performs
   in time $O(n^\omega \M(N)
  \log(N))$. When either $k>1$ or $k=1$ and $q^iA_0-\gamma_i\Id$ is
  invertible for~$0\le i< N$, this drops to
  $O(n^2\M(N)\log(N)+n^\omega N)$.
\end{Theo}

Our second algorithm behaves better with respect to~$N$, with cost
in~$O(\M(N))$ only, but it always involves polynomial matrix
multiplications. Since in many cases the divide-and-conquer approach
avoids these multiplications, the second algorithm becomes preferable
for rather large precisions.

In the differential case, when~$k=0$ and the characteristic is~0, the
algorithms in~\cite{BoChOlSaScSc07,BrKu78} compute an invertible matrix of power series
solution of the homogeneous equation by a Newton iteration and then
recover the solution using variation of the constant. In the more
general context we are considering here, such a matrix does not
exist. However, it turns out that an associated equation that can be
derived from~\eqref{eq:main} admits such a
solution. Section~\ref{sec:Newton} describes a variant of Newton's
iteration to solve it and obtains the following.
\begin{Theo}\label{theo:2}
  Assuming $A_0$ has good spectrum at precision~$N$, one can compute
  generators of the solution space of Eq.~\eqref{eq:main} at precision
  $N$ by a Newton-like iteration in time $O(n^\omega \M(
N)+n^\omega\log(n)N)$.
\end{Theo}
To the best of our knowledge, this is the first time such a low
complexity is reached for this problem. Without the good spectrum
assumption, however, we cannot guarantee that this algorithm succeeds,
let alone control its complexity.

%%%%%%%%%%%%%%%%%%%%%%%%%%%%%%%%%%%%%%%%%%%%%%%%%%%%%%%%%%%%
%%%%%%%%%%%%%%%%%%%%%%%%%%%%%%%%%%%%%%%%%%%%%%%%%%%%%%%%%%%%
%%%%%%%%%%%%%%%%%%%%%%%%%%%%%%%%%%%%%%%%%%%%%%%%%%%%%%%%%%%%

\section{Divide-and-Conquer}\label{sec:dac}

The classical approach to solving~\eqref{eq:main} is to proceed
term-by-term by coefficient extraction. Indeed, we can rewrite the
coefficient of degree $i$ in this equation as
\begin{equation}\label{eq:Ri}
\mR_i \mF_i = \Delta_i,  
\end{equation}
where $\Delta_i$ is a vector that can be computed from
$\mA$, $\mC$ and all previous $\mF_j$ (and whose actual expression
depends on $k$), and $\mR_i$ is as follows:
\[\begin{cases}
\mR_i = (q^i \mA_0-\gamma_i\Id)&\quad\text{if $k=1$}\\
\mR_i = q^i\mA_0&\quad\text{if $k>1$}.
\end{cases}\]
Ideally, we wish that each such system determines $\mF_i$ uniquely
that is, that $\mR_i$ be a unit. For $k=1$, this is the case when $i$
is not a root of the {\em indicial equation} $\det(q^i
\mA_0-\gamma_i\Id)=0$. For $k>1$, either this is the case for all $i$
(when $\mA_0$ is invertible) or for no $i$. In any case, we let
$\mathcal{R}$ be the set of indices $i \in\{0,\dots,N-1\}$ such that
$\det(\mR_i) = 0$; we write $\mathcal{R}=\{j_1 < \dots <j_r\}$, so
that $r=|\mathcal{R}|$.

Even when $\mathcal{R}$ is empty, so the solution is unique, this
approach takes quadratic time in $N$, as computing each individual
$\Delta_i$ takes linear time in $i$. To achieve quasi-linear time, we
split the resolution of Eq.~\eqref{eq:main} mod $x^N$ into two
half-sized instances of the problem; at the leaves of the recursion
tree, we end up having to solve the same Eq.~\eqref{eq:Ri}.

When $\mathcal{R}$ is empty, the algorithm is simple to state (and the
cost analysis simplifies; see the comments at the end of this
section). Otherwise, technicalities arise.  We treat the cases $i \in
\mathcal{R}$ separately, by adding placeholder parameters for all
corresponding coefficients of $\mF$ (this idea is already
in~\cite{BaBrPf10,BaPf99}; the algorithms in these references use a
finer classification when $k>1$, by means of a suitable extension of
the notion of indicial polynomial, but take quadratic time in $N$).

Let $\mathbf{f}_{1,1},\dots,\mathbf{f}_{n,r}$ be $nr$ new
indeterminates over $\K$ (below, all boldface letters denote
expressions involving these formal parameters). For $\rho=1,\dots,r$,
we define the vector~$\mathbf{F}_{j_\rho}$ with entries $\mathbf{f}_{1,\rho},\dots,\mathbf{f}_{n,\rho}$
and we denote by 
$\mathscr{L}$ the set of all vectors
\begin{equation*}
  \label{eq:bfF}
\mathbf{F}=\varphi_0 + \varphi_1 \mathbf{F}_{j_1}+ \cdots + \varphi_r
\mathbf{F}_{j_r},
\end{equation*}
with $\varphi_0$ in $\mathscr{M}_{n,1}(\K[x])$ and each $\varphi_\ell$
in $\mathscr{M}_{n}(\K[x])$ for $1 \le \ell\le r$.  
We also define $\mathscr{L}_i$ the subspace of vectors of the form
  $$\mathbf{F}=\varphi_0 + \varphi_1 \mathbf{F}_{j_1}+ \cdots +
  \varphi_{\mu(i)} \mathbf{F}_{\mu(i)},$$
where $\mu(i)$ is defined as the index of the largest
  element $j_\ell\in\mathcal{R}$ such that $j_\ell < i$; if no such
  element exist (for instance when $i=0$), we let $\mu(i)=0$.
A {\em specialization}
$S:\mathscr{L} \to \mathscr{M}_{n,1}(\K[x])$ is simply an evaluation
map defined by ${\bf f}_{i,\ell}\mapsto f_{i,\ell}$ for all
$i,\ell$, for some choice of $(f_{i,\ell})$ in $\K^{nr}$.

We extend
$\partial$ and $\sigma$ to such vectors, by letting
$\partial(\mathbf{f}_{i,\ell})=0$ and
$\sigma(\mathbf{f}_{i,\ell})=\mathbf{f}_{i,\ell}$ for all $i,\ell$, so
that we have, for $\mathbf{F}$ in $\mathscr{L}$
$$\partial(\mathbf{F})=\partial(\varphi_0) + \partial(\varphi_1)
\mathbf{F}_{j_1}+ \cdots + \partial(\varphi_r) \mathbf{F}_{j_r},$$ and
similarly for $\sigma(\mathbf{F})$.

\begin{algorithm}[t]
\caption{Recursive Divide-and-Conquer\label{algo:RDAC}}
\SetKwFunction{Algorithm}{$\RDAC$} \Algorithm{$A,\mathbf{C},i,N,k$}

\Input{$A\in\mathscr{M}_{n}(\K[[x]]), \mathbf{C}\in\mathscr{L}_i$,
$i\in\mathbb{N},N\in\mathbb{N}\setminus\{0\}, k\in\mathbb{N}\setminus\{0\}$}

\Output{$\mathbf{F}\in\mathscr{L}_{i+N}$}

\uIf{$N=1$}{
  \lIf{$k=1$}{$R_i:=q^i \mA_0-\gamma_i\Id$} \lElse{$R_i:=q^i\mA_0$}\\
  \lIf{$\det(R_i)=0$}{return $\mathbf{F}_i$}

  \lElse {return $-R_i^{-1} \mathbf{C}_0$}
}
\Else{
  $m:=\lceil N/2\rceil$\\
  $\mathbf{H}:=\RDAC(A,\mathbf{C},i,m,k)$\\
  $\mathbf{D}:=(\mathbf{C}-x^k\partial(\mathbf{H})+
  (q^iA-\gamma_i x^{k-1}\Id)\sigma(\mathbf{H})) {\rm~div~} x^m$\\
  $\mathbf{K}:=\RDAC(A,\mathbf{D},i+m,N-m,k)$\\
  return $\mathbf{H}+x^m \mathbf{K}$
}
\end{algorithm}

The main divide-and-conquer algorithm first computes $\mathbf{F}$ in
$\mathscr{L}$, by simply skipping all equations corresponding to
indices $i \in \mathcal{R}$; it is presented in
Algorithm~\ref{algo:DAC}. In a second step, we resolve the
indeterminacies by plain linear algebra.  For $i\ge 0$, and
$\mathbf{F},\mathbf{C}$ in $\mathscr{L}$, we write
$$ E(\mathbf{F},\mathbf{C},i) = x^k \partial(\mathbf{F}) - \Big( (q^i
A-\gamma_i x^{k-1} \Id) \sigma(\mathbf{F}) + \mathbf{C}\Big ).$$ In
particular, $E(\mathbf{F},\mathbf{C},0)$ is a parameterized form of
Eq.~\eqref{eq:main}.  The key to the divide-and-conquer approach is
to write $\mathbf{H}=\mathbf{F} \bmod x^m$,
$\mathbf{K}=\mathbf{F}{\rm~div~} x^m$ and ${\mathbf{D}}=(\mathbf{C}-
E(\mathbf{H},\mathbf{C},i)) {\rm~div~} x^m$.  Using the equalities
  \[x^k\partial (\mathbf{F})=x^k\partial (\mathbf{H})+x^{m+k}\partial (\mathbf{K})+\gamma_mx^{m+k-1}\sigma(\mathbf{K})\]
  and $\gamma_{i+m} = \gamma_m + q^m
  \gamma_i$, a quick computation shows that
  \begin{equation}\label{eq:E}
E(\mathbf{F},\mathbf{C},i) = \left( E(\mathbf{H},\mathbf{C},i) \bmod x^m \right )+ x^m E(\mathbf{K},{\mathbf{D}},i+m).
  \end{equation}

\begin{Lemma}\label{lemma:DAC}
Let $A$ be in $\mathscr{M}_{n}(\K[x])$ and $\mathbf{C}$ in $\mathscr{L}_i$, and let
  $\mathbf{F}=\RDAC(A,\mathbf{C},i,M,k)$ with $i+M\le N$. Then:
  \begin{enumerate}
  \item\label{ptE1} $\mathbf{F}$ is in $\mathscr{L}_{i+M}$;
  \item\label{ptE2} for $j\in \{0,\dots,M-1\}$ such that $i+j\not\in\mathcal{R}$, the equality $\mathrm{coeff}(E(\mathbf{F},\mathbf{C},i), x^j)=0$ holds;
  \item\label{ptE3} if $C$ and $F$ in $\mathscr{M}_{n,1}(\K[x])$ with $\deg F<M$ are such that $E(F,C,i)=0\bmod x^M$ and there exists a specialization $S:\mathscr{L}_{i} \to
      \mathscr{M}_{n,1}(\K[x])$ such that $C=S(\mathbf{C})$, 
      there exists a specialization $S':\mathscr{L}_{i+M} \to
      \mathscr{M}_{n,1}(\K[x])$ which extends $S$ and such that
      $F=S(\mathbf{F})$.
  \end{enumerate}
$\mathbf{F}$ is computed in time  $O((n^2+rn^\omega)
\M(M)\log(M)+n^\omega M)$.
\end{Lemma}
\begin{proof}
The proof is by induction on~$M$.

\noindent {\sc Proof of~\ref{ptE1}.} For $M=1$, we
distinguish two cases.  If $i\in\mathcal{R}$, say $i=j_\ell$, we
return $\mathbf{F}_i=\mathbf{F}_{j_\ell}$. In this case, 
$\mu(i+1)=\ell$, so our claim holds. If $i\not\in\mathcal{R}$,
because $\mathbf{C}_0\in\mathscr{L}_{i}$, the output is in
$\mathscr{L}_{i}$ as well. This proves the case $M=1$.

For $M>1$, we assume the claim to hold for all $(i,M')$, with $M'<M$.
By induction, $\mathbf{H}\in\mathscr{L}_{i+m}$ and $\mathbf{K}\in\mathscr{L}_{i+M}$. Thus,
${\mathbf{D}}\in\mathscr{L}_{i+m}$ and the conclusion follows.

\smallskip\noindent {\sc Proof of~\ref{ptE2}.} For~$M=1$, if $i\in\mathcal{R}$, the claim is trivially satisfied. Otherwise, we have to
verify that the constant term of $E(\mathbf{F},\mathbf{C},i)$ is
zero. In this case, the output $\mathbf{F}$ is reduced to its constant
term $\mathbf{F}_0$, and the constant term of
$E(\mathbf{F},\mathbf{C},i)$ is (up to sign)
$R_i\mathbf{F}_0+\mathbf{C}_0=0$, so we are done.

For $M>1$, we assume that the claim holds for all $(i,M')$, with
$M'<M$.  Take $j$ in $\{0,\dots,M-1\}$. If $j < m$, we have
$\mathrm{coeff}(E(\mathbf{F},\mathbf{C},i),
x^j)=\mathrm{coeff}(E(\mathbf{H},\mathbf{C},i), x^j)$; since $i+j
\notin \mathcal{R}$, this coefficient is zero by assumption.  If $m
\le j$, we have $\mathrm{coeff}(E(\mathbf{F},\mathbf{C},i),
x^j)=\mathrm{coeff}(E(\mathbf{K},{\mathbf{D}},i), x^{j-m})$. Now, $j+i
\notin \mathcal{R}$ implies that $(j-m)+(i+m) \notin \mathcal{R}$, and
$j-m < M-m$, so by induction this coefficient is zero as well.

\smallskip\noindent {\sc Proof of~\ref{ptE3}.} For $M=1$, if $i\in\mathcal{R}$, say $i=j_\ell$, we have
$\mathbf{F}=\mathbf{F}_{j_\ell}$, whereas $F$ has entries in $\K$;
this allows us to define $S'$. When $i\not\in\mathcal{R}$, we
have $F=S(\mathbf{F})$, so the claim holds as well. Thus, we are done
for $M=1$.

For $M>1$, we assume our claim for all $(i,M')$ with $M'<M$.  Write
$H=F \bmod x^m$, $K=F {\rm~div~} x^m$ and ${D}= (C-x^k\partial(H)+
(q^iA-\gamma_i x^{k-1}\Id)\sigma(H)) {\rm~div~}
x^m$. Then,~\eqref{eq:E} implies that $E(H,C,i) = 0\bmod x^m$ and
$E(K,{D},i+m)=0 \bmod x^{M-m}$. The induction assumption shows that
$H$ is a specialization of $\mathbf{H}$, say $H=S'(\mathbf{H})$ for
some $S':\mathscr{L}_{i+m} \to \mathscr{M}_{n,1}(\K[x])$ which extends
$S$. In particular, $D=S'({\mathbf{D}})$. The
induction assumption also implies that there exist an extension
$S'':\mathscr{L}_{i+m} \to \mathscr{M}_{n,1}(\K[x])$ of $S'$, and thus
of $S$, such that $K=S''(\mathbf{K})$. Then $F=S''(\mathbf{F})$, so we
are done.

\smallskip

For the complexity analysis, the most expensive part of the algorithm
is the computation of $\mathbf{D}$. At the inner recursion steps, the
bottleneck is the computation of $A \sigma(\mathbf{H})$, where
$\mathbf{H}$ has degree less than $M$ and $A$ can be truncated mod
$x^M$ (the higher degree terms have no influence in the subsequent
recursive calls).  Computing $\sigma(\mathbf{H})$ takes time
$O(N(n+rn^2))$ and the product is done in time
$O((n^2+rn^\omega)\M(M))$; recursion leads to a factor~$\log(M)$.  The
base cases use $O(M)$ matrix inversions of cost $O(n^\omega)$ and $O(M)$
multiplications, each of which takes time $O(r n^\omega)$.
\end{proof}
 
The second step of the algorithm is plain linear algebra: we know that
the output of the previous algorithm satisfies our main equation for
all indices $i \notin \mathcal{R}$, so we conclude by forcing the remaining ones to
zero.  
\begin{algorithm}[t]
\caption{Divide-and-Conquer\label{algo:DAC}}
\SetKwFunction{Algorithm}{$\DAC$} \Algorithm{$A,C,N,k$}

\Input{$A\in\mathscr{M}_{n}(\K[[x]]), C\in\mathscr{M}_{n,1}(\K[[x]])$,
$N\in\mathbb{N}\setminus\{0\}, k\in\mathbb{N}\setminus\{0\}$}

\Output{Generators of the solution space of $x^k \partial(F)=A \sigma(F)+C$ at precision $N$.}

$\mathbf{F} := \RDAC(A,C,0,N,k)$\\
\hfill ($\mathbf{F}$ has the form $\varphi_0 + \varphi_1 \mathbf{F}_{j_1}+ \cdots + \varphi_r
\mathbf{F}_{j_r}$)
$\mathbf{T}:=x^k\partial(\mathbf{F})-A\sigma(\mathbf{F})-C \bmod x^N$\\
$\Gamma:=(\mathbf{T}_i^{(j)},\ i \in \mathcal{R},\ j=1,\dots,n)$\\
$\Phi,\Delta:=\LinAlg(\Gamma=0)$\\
%% \lIf{$\Phi=\bot$}{return $\bot$}
%% \Else{
  $M := [ \varphi_1, \dots, \varphi_r ]$\\
  return $\varphi_0 + M \Phi, M \Delta$
%% }
\end{algorithm}

\begin{Prop}\label{prop:DAC}
  On input $A, C, N, k$ as specified in Algo\-rithm~\ref{algo:DAC}, this
  algorithm returns generators of the solution space
  of~\eqref{eq:main} mod $x^N$ in time $O((n^2+r n^\omega)
  \M(N)\log(N) + r^2 n^\omega N+ r^\omega n^\omega).$
\end{Prop}
\begin{proof}
  The first claim is a direct consequence of the construction above,
  combined with Lemma~\ref{lemma:DAC}.
% for $i=0$, $M=N$. 
For the cost estimate, we need
  to take into account the computation of $\mathbf{T}$, the linear
  system solving, and the final matrix products. The computation of
  $\mathbf{T}$ fits in the same cost as that of $\mathbf{D}$ in
  Algorithm~\ref{algo:RDAC}, so no new contribution comes from
  here. Solving the system $\Gamma=0$ takes time
  $O((rn)^\omega)$. Finally, the product $[ \varphi_1 \cdots \varphi_r ]
  \Delta$ involves an $n \times (rn)$ matrix with entries of degree
  $N$ and an $(rn) \times t$ constant matrix, with $t \le rn$;
  proceeding coefficient by coefficient, and using block matrix
  multiplication in size $n$, the cost is $O(r^2 n^\omega N)$.
\end{proof}
 
When all matrices $R_i$ are invertible, the situation becomes
considerably simpler: $r=0$, the solution space has dimension~0, there
is no need to introduce formal parameters, the cost drops to $O(n^2
\M(N)\log(N) + n^\omega N)$ for Lemma~\ref{lemma:DAC}, and
Proposition~\ref{prop:DAC} becomes irrelevant.

When $A_0$ has good spectrum at precision $N$, we may not be able to
ensure that $r=0$, but we always have $r \le 1$. Indeed, when $k=1$,
the good spectrum condition implies that for all $0\le i<N$ and
for $j\in\N$, the matrices $R_i$ and $R_j$
have disjoint spectra so that at most one of
them can be singular. For $k>1$, the good spectrum
condition implies that all $R_i$ are invertible, whence $r=0$. This proves
Thm.~\ref{theo:1}.

\smallskip\noindent{\bf Previous work.} As said above, Barkatou and
Pfl{\"u}gel~\cite{BaPf99}, then Barkatou, Broughton and
Pfl{\"u}gel~\cite{BaBrPf10}, already gave algorithms that solve such
equations term-by-term, introducing formal parameters to deal with
cases where the matrix $R_i$ is singular. These algorithms handle some
situations more finely than we do (e.g., the cases $k \ge 2$), but
take quadratic time; our algorithm can be seen as a
divide-and-conquer version of these results.

In the particular case $q\ne 1$, $n=1$ and $r=0$, another forerunner
to our approach is Brent and Traub's divide-and-conquer
algorithm~\cite{BrTr80}. That algorithm is analyzed for a more general
$\sigma$, of the form $\sigma(x)=xq(x)$, as such, they are more costly
than ours; when $q$ is constant, we essentially end up with the
approach presented here.

Let us finally mention van der Hoeven's paradigm of {\rm relaxed
  algorithms}~\cite{Hoeven2002,Hoeven09,Hoeven11}, which allows one to
solve systems such as~\eqref{eq:main} in a term-by-term fashion, but
in quasi-linear time. The cornerstone of this approach is fast {\em
  relaxed multiplication}, otherwise known as {\em online multiplication},
of power series. 

In~\cite{Hoeven2002,Hoeven2003}, van der Hoeven offers two relaxed
multiplication algorithms (the first one being similar to that
of~\cite{FischerStockmeyer1974}); both take time $O(\M(n)\log(n))$.
When $r=0$, this yields a complexity similar to Prop.~\ref{prop:DAC}
to solve Eq.~\eqref{eq:main}, but it is unknown to us how this
carries over to arbitrary $r$.

When $r=0$, both our divide-and-conquer approach and the relaxed one
can be seen as ``fast'' versions of quadratic time term-by-term
extraction algorithms. It should appear as no surprise that they
are related: as it turns out, at least in simple cases (with $k=1$ and
$n=1$),  using the relaxed multiplication algorithm
of~\cite{Hoeven2003} to solve Eq.~\eqref{eq:main} leads to doing
exactly the same operations as our divide-and-conquer method, without
any recursive call. We leave
the detailed analysis of these observations to future work.

For suitable ``nice'' base fields (e.g., for fields that support Fast
Fourier Transform), the relaxed multiplication algorithm
in~\cite{Hoeven2002} was improved in~\cite{Hoeven2007,Hoeven12}, by
means of a reduction of the $\log(n)$ overhead. This raises the
question whether such an improvement is available for divide-and
conquer techniques.

%%%%%%%%%%%%%%%%%%%%%%%%%%%%%%%%%%%%%%%%%%%%%%%%%%%%%%%%%%%%
%%%%%%%%%%%%%%%%%%%%%%%%%%%%%%%%%%%%%%%%%%%%%%%%%%%%%%%%%%%%
%%%%%%%%%%%%%%%%%%%%%%%%%%%%%%%%%%%%%%%%%%%%%%%%%%%%%%%%%%%%

\section{Newton Iteration}\label{sec:Newton}

\subsection{Gauge Transformation}\label{ssec:mF}
Let~$\mF$ be a solution of Eq.~\eqref{eq:main}.  To any invertible
matrix~$\mW\in\mathscr{M}_n(\K[x])$, we can associate the
matrix~$\mY=\mW^{-1}F\in\mathscr{M}_n({\K[[x]]})$. We are going to
choose~$\mW$ in such a way that~$\mY$ satisfies an equation simpler
than~\eqref{eq:main}. The heart of our contribution is the efficient
computation of such a~$\mW$.

\begin{Lemma}
Let $\mW\in\mathscr{M}_n(\K[x])$ be invertible in $\mathscr{M}_n(\K[[x]])$ and
let~$\mB\in\mathscr{M}_n(\K[x])$ be such that
\begin{equation}\label{eq:Wb}
\mB=\mW^{-1}(x^k \partial(\mW)-\mA \sigma(\mW)) \bmod x^N.
\end{equation}
Then $\mF$ in
  $\mathscr{M}_{n,1}(\K[x])$ satisfies
\begin{equation}\tag{\ref{eq:main}}
x^k\partial(\mF)=\mA\sigma(\mF) +\mC \bmod x^N
\end{equation} if and only if $\mY=\mW^{-1}\mF$ satisfies
\begin{equation}\label{eq:finaleq}
x^k\partial(\mY)=\mB\sigma(\mY)+ \mW^{-1}\mC  \bmod x^N.
\end{equation}
\end{Lemma}
\begin{proof}
Differentiating the equality $\mF=\mW \mY$ gives
$$x^k \partial(\mF) = x^k \partial(\mW) \sigma(\mY) + x^k \mW \partial(\mY).$$
Since 
$x^k\partial(\mW) = \mA \sigma(\mW)-\mW \mB \bmod x^N,$
we deduce
% $$x^k \partial(\mF) = (\mA \sigma(\mW)-\mW \mB)\sigma(\mY) + x^k\mW \partial(\mY) \bmod x^N.$$
% This implies that
$$x^k \partial(\mF)-\mA \sigma(\mF) -\mC = \mW (x^k \partial(\mY) -\mB
\sigma(\mY) -\mW^{-1}\mC) \bmod x^N.$$
Since $\mW$ is invertible, the conclusion follows.
\end{proof}

The systems~\eqref{eq:main} and~\eqref{eq:finaleq} are called
equivalent under the gauge transformation~$Y=WF$.
Solving~\eqref{eq:main} is thus reduced to finding a simple~$\mB$ such
that~\eqref{eq:finaleq} can be solved efficiently and such that the
equation
\begin{equation}\label{eq:W}
x^k \partial(\mW)=\mA \sigma(\mW)-\mW \mB \bmod x^N
\end{equation}
that we call \emph{associated} to~\eqref{eq:main} has an
\emph{invertible} matrix~$\mW$ solution that can be computed
efficiently too.

As a simple example, consider the differential case, with $k=1$. Under
the good spectrum assumption, it is customary to choose $\mB=\mA_0$,
the constant coefficient of $\mA$. In this case, the matrix $W$ of the
gauge transformation must satisfy $$x \mW' = \mA\mW-\mW \mA_0 \bmod
x^N.$$ It is straightforward to compute the coefficients of $\mW$ one
after the other, as they satisfy $\mW_0=\Id$ and, for $i>0$,
$$(\mA_0- i \Id)\mW_i -\mW_i \mA_0 = -\sum_{j < i}\mA_{i-j} \mW_j.$$ However, using
this formula leads to a quadratic running time in $N$.  The Newton
iteration presented in this section computes $\mW$ in quasi-linear
time.

%%%%%%%%%%%%%%%%%%%%%%%%%%%%%%%%%%%%%%%%%%%%%%%%%%%%%%%%%%%%

\subsection{Polynomial Coefficients}\label{ssec:polcoeffs}

Our approach consists in reducing efficiently the resolution
of~\eqref{eq:main} to that of an equivalent equation where the
matrix~$A$ of power series is replaced by a matrix~$B$ of polynomials
of low degree. This is interesting because the latter can be solved in
\emph{linear} complexity by extracting coefficients. This subsection
describes the resolution of the equation
\begin{equation}
  \label{eq:YP}x^k \partial(Y)=P \sigma(Y)+Q,
\end{equation}
where $P$ is a polynomial matrix of degree less than $k$.

\begin{algorithm}[t]
\caption{PolCoeffsDE\label{algo:PolCoeffsDE}}
\SetKwFunction{Algorithm}{$\PolCoeffsDE$}
\Algorithm{$P,Q,k,N$}

\Input{$P\in\mathscr{M}_n(\K[x])$ of degree less than~$k$, $Q\in\mathscr{M}_{n,1}(\K[[x]])$, $N\in\mathbb{N}\setminus\{0\}$, $k\in\mathbb{N}\setminus\{0\}$}

\Output{Generators of the solution space of  $x^k\partial(Y)=P\sigma(Y)+Q$ at precision $N$.} 

\For{$i=0,\dots,N-1$}{
  $C:=Q_i+(P_1q^{i-1}Y_{i-1}+\dots+P_{k-1}q^{i-k+1}Y_{i-k+1})$\\
  \uIf{$k=1$}{$Y_i,M_i:=\LinAlg((\gamma_i\Id-q^iP_0)X=C)$}
  \Else{$Y_i,M_i:=\LinAlg(-q^iP_0X=C-\gamma_{i-k+1}Y_{i-k+1})$}}
return $Y_0+\dots+Y_{N-1}x^{N-1}$, $[M_0~M_1x \cdots M_{N-1}x^{N-1}]$
\end{algorithm}

\begin{Lemma}\label{lemma:PolCoeffsDE}
  Suppose that $P_0$ has good spectrum at precision $N$. Then
  Algorithm~\ref{algo:PolCoeffsDE} computes generators of the
  solution space of Eq.~\eqref{eq:YP} at precision $N$ in time
  $O(n^\omega N)$, with $M\in\mathscr{M}_{n,t}(\K)$ for some $t \le n$.
\end{Lemma}
\begin{proof}
  Extracting the coefficient of~$x^i$ in Eq.~\eqref{eq:YP} gives
  \[\gamma_{i-k+1}Y_{i-k+1}=q^iP_0Y_i+\dots+q^{i-k+1}P_{k-1}Y_{i-k+1}+Q_i.\]
  In any case, the equation to be solved is as indicated in the
  algorithm. For $k=1$, we actually have $C=Q_i$ for all $i$, so all
  these systems are independent. For $k>1$, the good spectrum
  condition ensures that the linear system has full rank for all
  values of $i$, so all $M_i$ are empty. For each~$i$, computing~$C$
  and solving for~$Y_i$ is performed in~$O(n^\omega)$ operations,
  whence the announced complexity.
\end{proof}

%%%%%%%%%%%%%%%%%%%%%%%%%%%%%%%%%%%%%%%%%%%%%%%%%%%%%%%%%%%%

\subsection{Computing the Associated Equation}\label{ssec:mB}

Given~$A\in\mathscr{M}_n(\K[[x]])$, we are looking for a matrix~$B$
with polynomial entries of degree less than~$k$ such that the
associated equation~\eqref{eq:W}, which does not depend on the
non-homogeneous term~$C$, has an invertible matrix
solution.

In this article, we content ourselves with a simple version of the
associated equation where we choose~$B$ in such a way
that~\eqref{eq:W} has an invertible solution~$V\bmod x^k$; thus, $V$
and $B$ must satisfy $\mA \sigma(\mV)=\mV \mB \bmod x^k$. The
invertible matrix $V$ is then lifted at higher precision by Newton
iteration (Algorithm~\ref{algo:NewtonAE}) under regularity conditions
that depend on the spectrum of~$A_0$. Other cases can be reduced to
this setting by the polynomial gauge transformations that are used in the
computation of formal
solutions~\cite{BaBrPf10,Wasow65}.

%We determine simultaneously~$B$ and~$V$.
When $k=1$ or $q\ne 1$, the
choice
\[B=A\bmod x^k,\quad V=\Id\]
solves our constraints and is sufficient to solve the associated
equation. When~$q=1,k>1$ (in particular when the point~0 is an irregular singular
point of the equation), this is not be the case anymore. In that
case, we use a known technique called the {\em splitting lemma} to
prepare our equation. See for instance~\cite[Ch.~3.2]{Balser00} and~\cite{BaBrPf10} for
details and generalizations.

\begin{Lemma}[Splitting Lemma]\label{splittinglemma}
Suppose that $k>1$, that $|\Spec A_0|=n$ and that $\Spec A_0 \subset
\K$. Then one can compute in time $O(n^\omega)$ matrices $\mV$ and
$\mB$ of degree less than $k$ in $\mathscr{M}_n(\K[x])$ such that the
following holds: 
$\mV_0$ is invertible;
$\mB$ is diagonal;
$\mA\mV=\mV \mB \bmod x^k$.
  % \begin{itemize}
  % \item $\mV_0$ is invertible;
  % \item $\mB$ is diagonal;
  % \item $\mA\mV=\mV \mB \bmod x^k$.
  % \end{itemize}
\end{Lemma}
\begin{proof}
We can assume that $\mA_0$ is diagonal: if not, we let $\mP$ be in
$\mathscr{M}_n(\K)$ such that $\mD=\mP^{-1}\mA\mP$ has a diagonal
constant term; we find $\mV$ using $\mD$ instead of $\mA$, and replace
$\mV$ by $\mP \mV$. Computing $\mP$ and $\mP\mV$ takes time
$O(n^\omega)$, since as per convention, $k$ is considered constant in
the cost analyses.

Then, we take $\mB_0=\mA_0$ and $\mV_0=\Id$.  For $i>0$, we have to
solve $\mA_0 \mV_i-\mV_i \mA_0 -\mB_i = \Delta_i,$ where $\Delta_i$
can be computed from $\mA_1,\dots,\mA_i$ and $\mB_1,\dots,\mB_{i-1}$
%by matrix products, 
in time $O(n^\omega)$. We set the
diagonal of $V_i$ to 0. Since~$A_0$ is diagonal, the diagonal~$B_i$ is
then equal to the diagonal of $\Delta_i$, up to sign. Then the entry
$(\ell,m)$ in our equation reads
$(r_\ell-r_m)V_i^{(\ell,m)}=\Delta_i^{(\ell,m)}$, with
$r_1,\dots,r_n$ the (distinct) eigenvalues of $A_0$. 
This can always be solved, in a unique way. The total time is
$O(n^\omega)$.
\end{proof}

%%%%%%%%%%%%%%%%%%%%%%%%%%%%%%%%%%%%%%%%%%%%%%%%%%%%%%%%%%%%

\subsection{Solving the Associated Equation}\label{ssec:mW}

Once $\mB$ and $\mV$ are determined as in \S\ref{ssec:mB}, we compute
a matrix $\mW$ that satisfies the associated equation~\eqref{eq:W};
this  eventually allows us to reduce~\eqref{eq:main} to an equation
with polynomial coefficients.  This computation of~$\mW$ is performed
efficiently using a suitable version of Newton iteration for
Eq.~\eqref{eq:W}; it computes a sequence of matrices whose precision
is roughly doubled at each stage.  This is described in
Algorithm~\ref{algo:NewtonAE}; our main result in this section is the
following.

\begin{Prop}\label{prop:W}
  Suppose that $A_0$ has good spectrum at precision $N$.  Then, given
  a solution of the associated equation mod $x^k$, invertible
  in~$\mathscr{M}_n(\K[[x]])$, Algorithm~\ref{algo:NewtonAE} computes
  a solution of that equation  $\bmod x^N$, also invertible
  in $\mathscr{M}_n(\K[[x]])$, in time $O(n^\omega \M(N)+n^\omega
  \log(n) N)$.
\end{Prop}

Before proving this result, we show how to solve yet another type of
equations that appear in an intermediate step:
\begin{equation}\label{eq:WB}
x^k \partial(\mU) = \mB \sigma(\mU)- \mU\mB + \Gamma \bmod x^N,
\end{equation}
where all matrices involved have size $n\times n$, with $\Gamma = 0 \bmod x^m$. This is dealt with by
Algorithm~\ref{algo:DiffSylvester} when $k = 1$ or $q\ne 1$ and Algorithm~\ref{algo:DiffSylvesterDiff} otherwise.

For Algorithm~\ref{algo:DiffSylvester}, remember that $B=A \bmod~x^k$.
The algorithm uses a routine~$\Sylvester$ solving {\em Sylvester
  equations}. Given matrices $Y,V,Z$ in $\mathscr{M}_n(\K)$, we are
looking for $X$ in $\mathscr{M}_{n}(\K)$ such that $Y X-XV=Z.$ When
$(Y,V)$ have disjoint spectra, this system admits a unique solution,
which can be computed $O(n^\omega\log(n))$ operations
in~$\K$~\cite{Kirrinnis01}.

\begin{algorithm}[t]
\caption{Solving Eq.~\eqref{eq:WB} when $k=1$ or $q\neq1$\label{algo:DiffSylvester}}
\SetKwFunction{Algorithm}{$\DiffSylvester$}
\Algorithm{$\Gamma,m,N$}

\Input{$\Gamma\in x^m\mathscr{M}_{n}(\K[[x]]),m\in\mathbb{N}\setminus\{0\},N\in\mathbb{N}\setminus\{0\}$}

\Output{$U\in x^{m-k}\mathscr{M}_{n}(\K[x])$ solution of~\eqref{eq:WB}.} 

\For{$i=m,\dots,N-1$}{
    $C:=(B_1q^{i-1}U_{i-1}+\dots+B_{k-1}q^{i-k+1}U_{i-k+1})$\\
    $\qquad-(U_{i-1}B_1+\dots+U_{i-k+1}B_{k-1})+\Gamma_i$\\
    \uIf{$k=1$}{$U_i:=\Sylvester(XB_0+(\gamma_i\Id-q^iB_0)X=C)$}
    \Else{$U_i:=\Sylvester(XB_0-q^iB_0X=$\\
      $\qquad\qquad C-\gamma_{i-k+1}U_{i-k+1})$}
  }
return $U_mx^m+\dots+U_{N-1}x^{N-1}$
\end{algorithm}

\begin{Lemma}\label{coro1}
  Suppose that~$k=1$ or $q\neq1$ and that $A_0$ has good
  spectrum at precision $N$. 
  If $\Gamma= \mzero \bmod x^m$, with $k\le m<N$,
  then Algorithm~\ref{algo:DiffSylvester} computes a solution $\mU$ to
  Eq.~\eqref{eq:WB} that satisfies $\mU=\mzero \bmod x^{m-k+1}$ in
  time $O(n^\omega \log(n) N)$.
\end{Lemma}
\begin{proof}
Extracting the coefficient of~$x^i$ in~\eqref{eq:WB} gives
\[\gamma_{i-k+1}\mU_{i-k+1}=q^iB_0\mU_i-\mU_iB_0+C,\]
with~$C$ as defined in Algorithm~\ref{algo:DiffSylvester}.  In both
cases $k=1$ and $k>1$, this gives a Sylvester equation for each~$\mU_i$,
of the form given in the algorithm. Since $B_0=A_0$, the spectrum
assumption on $A_0$ implies that these equations all have a unique
solution. Since $\Gamma$ is $\mzero\bmod x^{m}$, so is $\mU$ (so
we can start the loop at index $m$). The
total running time is $O(n^\omega \log(n)N)$ operations
in~$\K$. \end{proof}

\begin{algorithm}[t]
\caption{Solving Eq.~\eqref{eq:WB} when $k>1$ and $q=1$\label{algo:DiffSylvesterDiff}}
\SetKwFunction{Algorithm}{$\mathsf{DiffSylvesterDifferential}$}

\Algorithm{$\Gamma,m,N$}

\Input{$\Gamma\in x^m\mathscr{M}_{n}(\K[[x]]),m\in\mathbb{N}\setminus\{0\},N\in\mathbb{N}\setminus\{0\}$}
\Output{$\mU\in x^{m-k}\mathscr{M}_{n}(\K[x])$ solution of~\eqref{eq:WB}.}
\For{$i=1,\dots,n$}{
	\For{$j=1,\dots,n$}{
		\lIf{$i=j$}{$\mU^{(i,i)}:=x^k\int(x^{-k}\Gamma^{(i,i)}) \bmod x^N$}

		\Else{$\mU^{(i,j)}\!\!:=\PolCoeffsDE(B^{(i,i)}\!\!-B^{(j,j)}\!\!,\Gamma^{(i,j)}\!\!,k,N)$}
	}}
	return $\mU$
\end{algorithm}

\medskip

This approach fails in the differential case ($q=1$) when $k>1$, since
then the Sylvester systems are all
singular. Algorithm~\ref{algo:DiffSylvesterDiff} deals with this
issue, using the fact that in this case, $B$ is diagonal, and
satisfies the conditions of Lemma~\ref{splittinglemma}.
\begin{Lemma} \label{coro2}
  Suppose that~$k>1$,~$q=1$ and that $A_0$ has good spectrum at
  precision $N$.  If $\Gamma= \mzero \bmod x^m$, with $k\le m<N$, then Algorithm~\ref{algo:DiffSylvesterDiff}
  computes a solution~$\mU$ to Eq.~\eqref{eq:WB} that
  satisfies~$\mU=0\bmod x^{m-k+1}$ in time~$O(n^2 N)$.
\end{Lemma}
\begin{proof}
Since~$B$ is diagonal, the $(i,j)$th entry of~\eqref{eq:WB} is
\[x^k \partial(\mU^{(i,j)}) = (B^{(i,i)}-B^{(j,j)}) \mU^{(i,j)} + \Gamma^{(i,j)} \bmod x^N.\]
When $i=j$, $B^{(i,i)}-B^{(j,j)}$ vanishes. After dividing by $x^k$,
we simply have to compute an integral, which is feasible under the
good spectrum assumption (we have to divide by the non-zero
$\gamma_1=1,\dots,\gamma_{N-k}=N-k$). When $i\ne
j$, the conditions ensure that Lemma~\ref{lemma:PolCoeffsDE} applies
(and since $k>1$, the solution is unique, as pointed out in %the proof of that lemma).
its proof).
\end{proof}

We now prove the correctness of Algorithm~\ref{algo:NewtonAE} for
Newton iteration. Instead of doubling the precision at each step,
there is a slight loss of $k-1$.
\begin{algorithm}[t]
\caption{Newton iteration for Eq.~\eqref{eq:W}\label{algo:NewtonAE}}
\SetKwFunction{Algorithm}{$\NewtonAE$}
\Algorithm{$V,N$}

\Input{$V\in\mathscr{M}_{n}(\K[x])$ solution of~\eqref{eq:W} $\bmod
  x^k$ invertible in~$\mathscr{M}_n(\K[[x]])$,
  $N\in\mathbb{N}\setminus\{0\}$}

\Output{$W\in\mathscr{M}_{n}(\K[x])$ solution of~\eqref{eq:W} $\bmod
  x^N$ invertible in~$\mathscr{M}_n(\K[[x]])$, with $W=V\bmod x^k$}

\lIf{$N\le k$}{return $V$}

\Else{
	$m:=\lceil{\frac{N+k-1}{2}}\rceil$

	$H:=\NewtonAE(V,m)$

	${R}:=x^k\partial(H)-A\sigma(H)+HB$

	\uIf{$k=1$ {\em or} $q\neq1$}{$U:=\DiffSylvester(-H^{-1}R,m,N)$}
	\Else{$U:=\DiffSylvesterDifferential(-H^{-1}R,m,N)$}

	return $H+HU$
}
\end{algorithm}

\begin{Lemma}\label{lemma:mu}
  Let $m\ge k$ and let $H\in\mathscr{M}_n(\K[x])$ be 	invertible in
	  $\mathscr{M}_{n}(\K[[x]])$ and satisfy~\eqref{eq:W}  $\bmod x^m$. Let $N$ be such that $m \le N \le
  2m-k+1$. Let $R$ and $U$ be as in Algorithm~\ref{algo:NewtonAE} and
  suppose that $A_0$ has good spectrum at precision $N$.

  Then $H+HU$ is invertible in $\mathscr{M}_{n}(\K[[x]])$ and
  satisfies the associated equation  $\bmod x^{N}$. Given $H$, $U$ can
  be computed in time $O(n^\omega \M(N)+n^\omega\log(n)N)$.
\end{Lemma}
\begin{proof}
%Since~$H$ satisfies~\eqref{eq:W} $\bmod~x^m$, 
By hypothesis,
$R = 0 \bmod
x^m$. Then
\begin{align*}
x^k\partial(H&+HU)-A\sigma(H+HU)+(H+HU)B\\
&=(x^k\partial(H)-A\sigma(H)+HB)(\Id+\sigma(U))\\
&\qquad+H(x^k\partial(U)+UB-B\sigma(U))\\
&=R(\Id+\sigma(U))-R\bmod x^N=R\sigma(U)\bmod x^N.
\end{align*}
Using either Lemma~\ref{coro1} or Lemma~\ref{coro2}, $U=\mzero \bmod
x^{m-k+1}$, so $\sigma(U)=0\bmod x^{m-k+1}$. Thus, the latter
expression is 0, since~$2m-k+1\ge N$. Finally, since $HU=\mzero \bmod
x^{m-k+1}$, and $m\ge k$, $H+HU$ remains invertible in
$\mathscr{M}_n(\K[[x]])$. The various matrix products and inversions
take a total number of $O(n^\omega \M(N))$ operations in~$\K$ (using
Newton iteration to invert $H$). Adding the cost of Lemma~\ref{coro1},
resp. Lemma~\ref{coro2}, we get the announced complexity.
\end{proof}

We can now prove Proposition~\ref{prop:W}.  Correctness is obvious by
repeated applications of the previous lemma. The cost ~$C(N)$ of the
computation up to precision~$N$ satisfies
\[C(N)=C(m)+O(n^\omega\M(N)+n^\omega\log nN),\quad N>k.\]
Using the super-additivity properties of the function $\M$ as
in~\cite[Ch.~9]{GaGe99}, we obtain the claimed complexity. 

%%%%%%%%%%%%%%%%%%%%%%%%%%%%%%%%%%%%%%%%%%%%%%%%%%%%%%%%%%%%

We can now conclude the proof of Thm.~\ref{theo:2}. In order to solve
Equation~\eqref{eq:main}, we first determine $B$ and $V$ as
in~\S\ref{ssec:mB}; the cost will be negligible. Then, we use
Proposition~\ref{prop:W} to compute a matrix $\mW$ that
satisfies~\eqref{eq:W} $\bmod x^N$. Given $\mC$ in
$\mathscr{M}_{n,1}(\K[[x]])$, we next compute $\Gamma=\mW^{-1} \mC \bmod
x^N$. By the previous lemma, we conclude by solving
\[x^k\partial(Y)=B\sigma(Y)+\Gamma\bmod x^N.\]
Lemma~\ref{lemma:PolCoeffsDE} gives us generators of the solution
space of this equation $\bmod x^N$. If it is
inconsistent, we infer that Eq.~\eqref{eq:main} is. Else, from
the generators $(Y,M)$ obtained in Lemma~\ref{lemma:PolCoeffsDE}, we
deduce that $(WY,WM) \bmod x^N$ is a generator of the solution space
of Eq.~\eqref{eq:main} $\bmod x^N$. Since the matrix $M$ has few
columns (at most $n$), the cost of all these computations is dominated
by that of Proposition~\ref{prop:W}, as reported in
Thm.~\ref{theo:2}.

%%%%%%%%%%%%%%%%%%%%%%%%%%%%%%%%%%%%%%%%%%%%%%%%%%%%%%%%%%%%
%%%%%%%%%%%%%%%%%%%%%%%%%%%%%%%%%%%%%%%%%%%%%%%%%%%%%%%%%%%%
%%%%%%%%%%%%%%%%%%%%%%%%%%%%%%%%%%%%%%%%%%%%%%%%%%%%%%%%%%%%

\section{Implementation}

We implemented the divide-and-conquer and Newton iteration algorithms,
as well as a quadratic time algorithm, on top of
NTL~5.5.2~\cite{NTL}. In our experiments, the base field is
$\K=\mathbb{Z}/p \mathbb{Z}$, with $p$ a 28 bit prime; the systems
were drawn at random. Timings are in seconds, averaged over 50 runs;
they are obtained on a single core of a 2 GHz Intel Core 2.

Our implementation uses NTL's built-in \texttt{zz\_pX} polynomial
arithmetic, that is, works with ``small'' prime fields (of size about
$2^{30}$ over 32 bit machines, and $2^{50}$ over 64 bits machines).
For this data type, NTL's polynomial arithmetic uses a combination of
naive, Karatsuba and FFT arithmetic.

There is no built-in NTL type for polynomial matrices, but a simple
mechanism to write one. Our polynomial matrix product is naive, of
cubic cost. For small sizes such as $n=2$ or $n=3$, this is
sufficient; for larger $n$, one should employ improved schemes (such
as Waksman's~\cite{Waksman70}, see also~\cite{DrIsSc11}) or
evaluation-interpolation techniques~\cite{BoSc05}.

Our implementation follows the descriptions given above, up to a few
optimizations for algorithm $\NewtonAE$ (which are all classical
in the context of Newton iteration). For instance, the inverse of
$H$ should not be recomputed at every step, but simply updated; some
products can be computed at a lower precision than it appears (such as
$H^{-1} R$, where $R$ is known to have a high valuation).

In Fig.~\ref{fig:1d}, we give timings for the scalar case, with $k=1$
and $q \ne 1$. Clearly, the quadratic algorithm is outperformed for
almost all values of $N$; Newton iteration performs better than the
divide-and-conquer approach, and both display a subquadratic behavior.
Fig.~\ref{fig:2d} gives timings when $n$ varies, taking $k=1$ and
$q\neq 1$ as before. For larger values of $n$, the divide-and-conquer
approach become much better for this range of values of~$N$, since it
avoids costly polynomial matrix multiplication (see
Thm.~\ref{theo:1}).

\begin{figure}[t]
\centerline{\includegraphics[width=8cm]{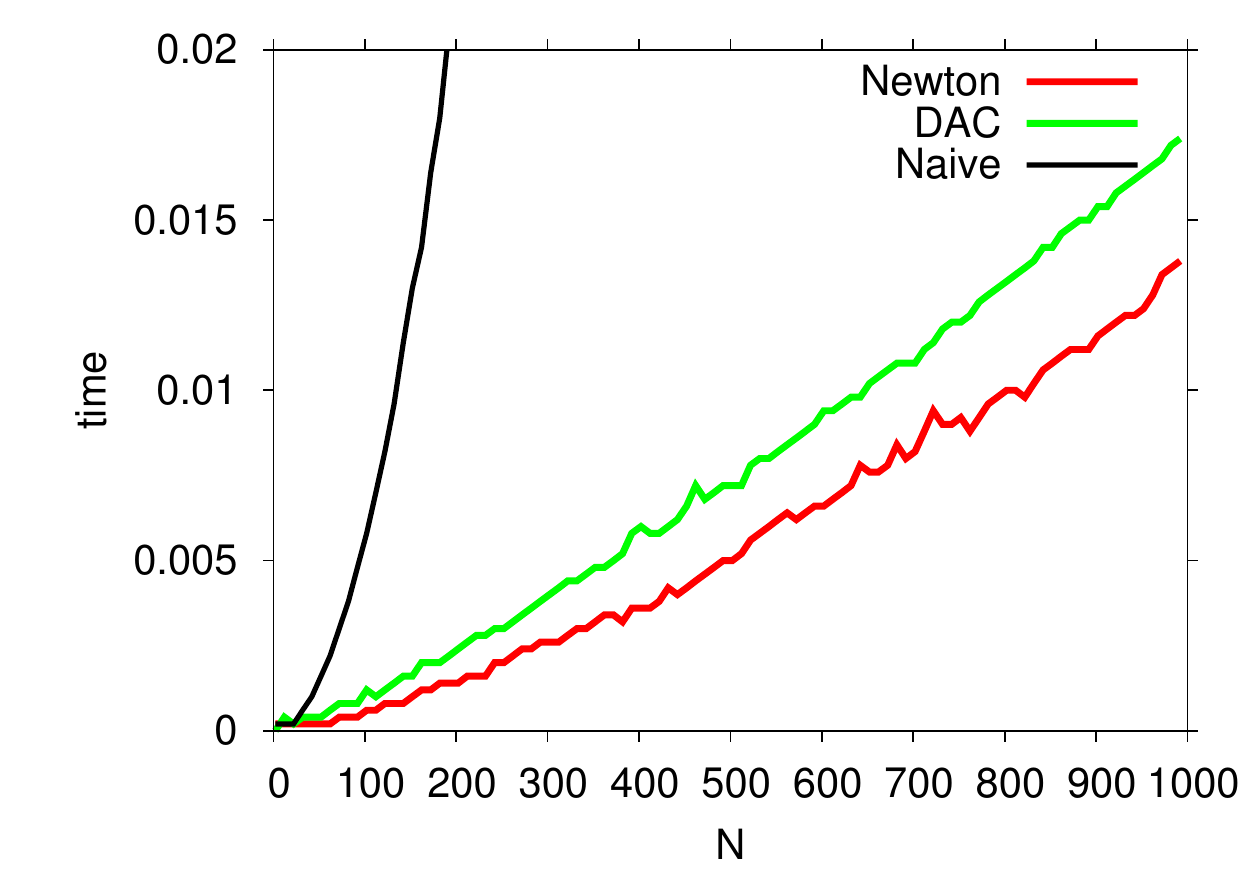}}
\vspace{-0.5cm}
\caption{Timings with $n=1$, $k=1$, $q \ne 1$}
\label{fig:1d}
\centerline{\includegraphics[width=8cm]{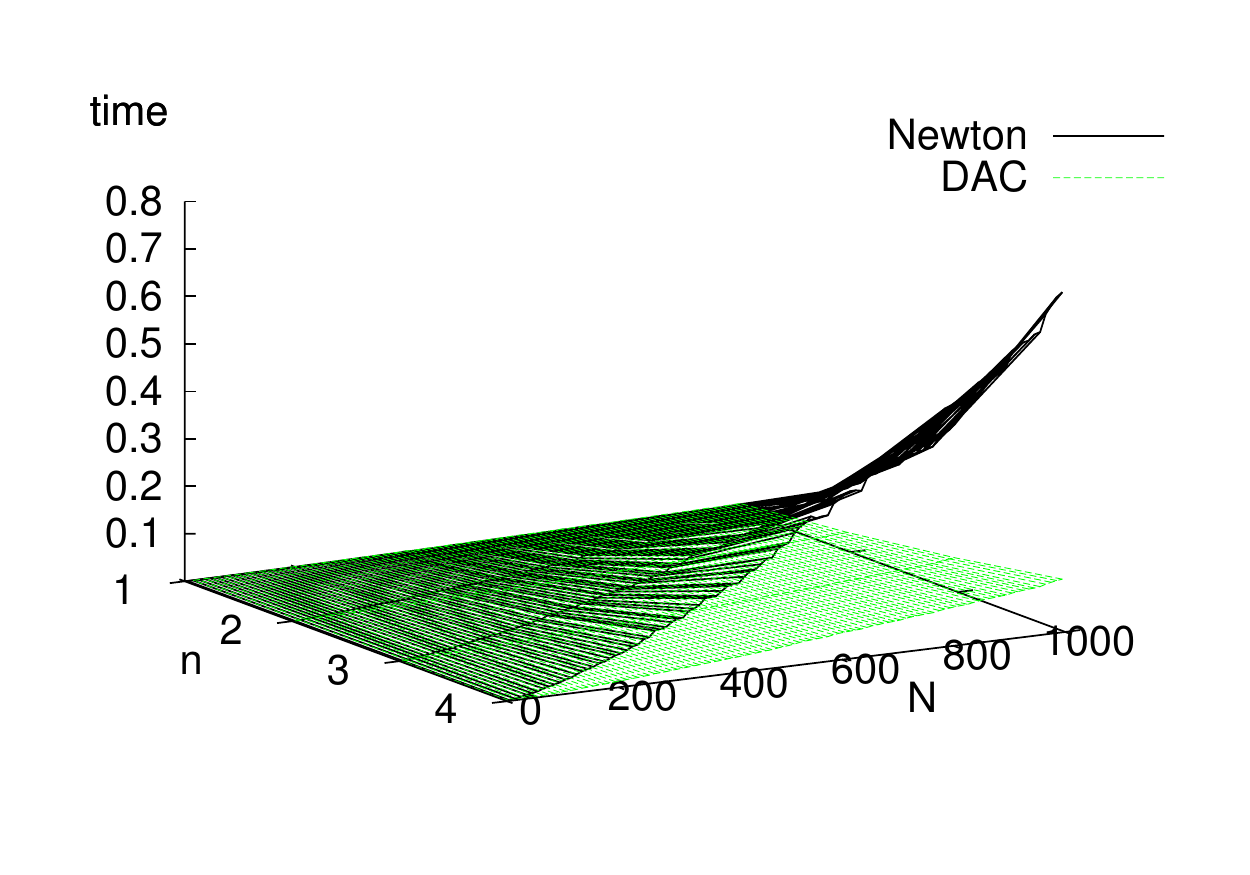}}
\vspace{-1cm}
\caption{Timings with $k=1$, $q \ne 1$}
\label{fig:2d}
\end{figure}

Finally, Table~\ref{tab:1} gives timings obtained for $k=3$, for
larger values of $n$ (in this case, a plot of the results would be
less readable, due to the large gap between the divide-and-conquer
approach and Newton iteration, in favor of the former); DAC stands for
``divide-and-conquer''. In all cases, the experimental results
confirm to a very good extent the theoretical cost analyses.

\begin{table}[t]
  \centerline{
    \begin{tabular}{|c|c||c|c|c|c|c|}
      \hline
      \multicolumn{2}{|c|}{Newton}&\multicolumn{4}{c|}{$n$}\\
      \cline{3-6}
      \multicolumn{2}{|c|}{}&5&9&13&17\\
      \hline
      \multirow{4}{*}{$N$}
                          & 50 & 0.01 & 0.11 & 0.32 & 0.72\\
      \cline{2-6}         & 250 & 0.22 & 1.2 & 3.7 & 8.1\\
      \cline{2-6}         & 450 & 0.50 & 2.8 & 8.3 & 18\\
      \cline{2-6}         & 650 & 0.93 & 5.1 & 16 & 34\\
      \hline
    \end{tabular}
}

  \centerline{
    \begin{tabular}{|c|c||c|c|c|c|c|}
      \hline
      \multicolumn{2}{|c|}{DAC}&\multicolumn{4}{c|}{$n$}\\
      \cline{3-6}
      \multicolumn{2}{|c|}{}&5&9&13&17\\
      \hline
      \multirow{4}{*}{$N$}
                          & 50 & 0.01 & 0.01 & 0.02 & 0.04\\
      \cline{2-6}         & 250 & 0.03 & 0.07 & 0.15 & 0.25\\
      \cline{2-6}         & 450 & 0.06 & 0.16 & 0.32 & 0.52\\
      \cline{2-6}         & 650 & 0.10 & 0.27 & 0.53 & 0.88\\
      \hline
    \end{tabular}
  }
\caption{Timings with $k=3$, $q \ne 1$}
\label{tab:1}
\end{table}

\scriptsize
\bibliographystyle{abbrv}
%\bibliography{main}

\end{document}